\documentclass{ptapap}

\author{Swayamtrupta Panda}[CFT,CAMK]
\author{Bo{\.z}ena Czerny}[CFT,CAMK]
\author{Conor Wildy}[CFT]
\author{Marzena {\'S}niegowska}[CFT,UW]
\affil[CAMK]{Nicolaus Copernicus Astronomical Center, ul. Bartycka 18, 00--716 Warsaw, Poland}
\affil[CFT]{Center for Theoretical Physics, Al. Lotników 32/46, 02--668 Warsaw, Poland}
\affil[UW]{Warsaw University Observatory, Al. Ujazdowskie 4, 00-478 Warsaw, Poland}

\title{What drives the Quasar Main Sequence?}
\begin{document}

\maketitle

\begin{abstract}
Eigenvector 1 (EV1) was found to be the dominant component behind the significant correlations for the measured parameters in quasar spectra (\citealt{bg92}). The parameter R$_{\mathrm{FeII}}$, which strongly correlates to the EV1, is the ${\mathrm{FeII}}$ strength, defined to be the ratio of the equivalent width of ${\mathrm{FeII}}$ to the equivalent width of ${\mathrm{H\beta}}$. This allows to construct a quasar main sequence analogous to the stellar properties driven HR diagram (\citealt{sul01}). We try to find the main driver behind the EV1 among the basic (theoretically motivated) parameters of an active nucleus (Eddington ratio, black hole mass, accretion rate, spin, and viewing angle). Based on theoretical modeling using the photoionization code CLOUDY (\citealt{f13}), we test the hypothesis that the physical driver of EV1 is the maximum of the accretion disk temperature ($\mathrm{T_{BBB}}$), reflected in the shape of the spectral energy distribution (SED). We have assumed that both H$\mathrm{\beta}$ and Fe${\mathrm{II}}$ emission come from the Broad Line Region represented as a constant density cloud in a plane-parallel geometry. We test the effect of changing Eddington ratio on the $\mathrm{R_{FeII} - T_{BBB}}$ trends with varying mean hydrogen densities. We also test the effect of adding microturbulence that affect the line intensities on the overall $\mathrm{R_{FeII} - T_{BBB}}$ picture. 
\end{abstract}

\section{Introduction}
Its been over two decades searching for a physically motivated parameter guiding the main sequence for quasars (\citealt{bg92,sul00,sul02,sul07,yip04,sh14,sun15}). Using the principal component analysis, it has been found that this sequence is driven mostly by the Eddington ratio (\citealt{bg92,sul00,sh14}) but also with the additional effect of the black hole mass, viewing angle and the intrinsic absorption (\citealt{sh14,sul00,kura09}). Eigenvector 1 (EV1) is dominated by the anti-correlation between the Fe${\mathrm{II}}$ optical emission and [OIII] line which itself accounts for 30\% of the total variance. The parameter R$_{\mathrm{FeII}}$, which strongly correlates to the EV1, is the ${\mathrm{FeII}}$ strength, defined to be the ratio of the equivalent width of ${\mathrm{FeII}}$ to the equivalent width of ${\mathrm{H\beta}}$. We postulate that the true driver behind the R$_{\mathrm{FeII}}$ is the maximum of the temperature in a multicolor accretion disk which is also the basic parameter determining the broad band shape of the quasar continuum emission. The prescription in detail is provided in \cite{panda17a} and \cite{panda17b}.

\section{Results and Discussions}
In \citet{panda17b}, we found that by changing the basic assumption from ``\textit{considering a constant bolometric luminosity}'' (Method 1) to ``\textit{considering a constant Eddington ratio}'' (Method 2), along with the presence of the hard X-ray power law and using the observational relation between UV and X-ray luminosities from \cite{lusso17} to determine the broad band spectral index, $\mathrm{\alpha_{ox}}$, changed the behaviour of the trend between $\mathrm{R_{FeII} - T_{BBB}}$ from monotonically declining to monotonically rising in the considered range of maximum of the disk temperatures [$5\times 10^4\;\mathrm{K}, 5\times 10^5\; \mathrm{K}$]. Additionally we found that, with increase in mean hydrogen density ($\mathrm{n_H}$) of the single-cloud (going from $10^{10}\;\mathrm{cm^{-3}}$ to $10^{11}\;\mathrm{cm^{-3}}$), the FeII strength i.e. $\mathrm{R_{FeII}}$ goes down by a factor 2. In other words, the FeII emission gets suppressed as we increase the mean hydrogen density. The dependence on the change of the Eddington ratio is not as significant as thought. Fig. 1(a,b) shows the effect of changing the Eddington ratio on the $\mathrm{R_{FeII} - T_{BBB}}$. In Fig. 1(a), we show the $\mathrm{R_{FeII} - T_{BBB}}$ trends by changing the Eddington ratio from 0.1 to 1. For both these cases, we vary the mean hydrogen density ($\mathrm{n_H}$) from $\mathrm{10^{10}\; cm^{-3}}$ to $\mathrm{10^{11}\; cm^{-3}}$. At lower temperatures ($\mathrm{T_{BBB} < 2.24 \times 10^4\;K}$) we see again a rise in the value of $\mathrm{R_{FeII}}$ for both cases. This rise is due to the rise in the value of the $\mathrm{\alpha_{ox}}$ for $\mathrm{T_{BBB} < 2.24 \times 10^4\;K}$ as illustrated in Fig. 2(a).

\par  
\begin{figure}[ht]
\begin{minipage}[b]{0.45\linewidth}
\centering
\includegraphics[width=0.75\textwidth, angle=270]{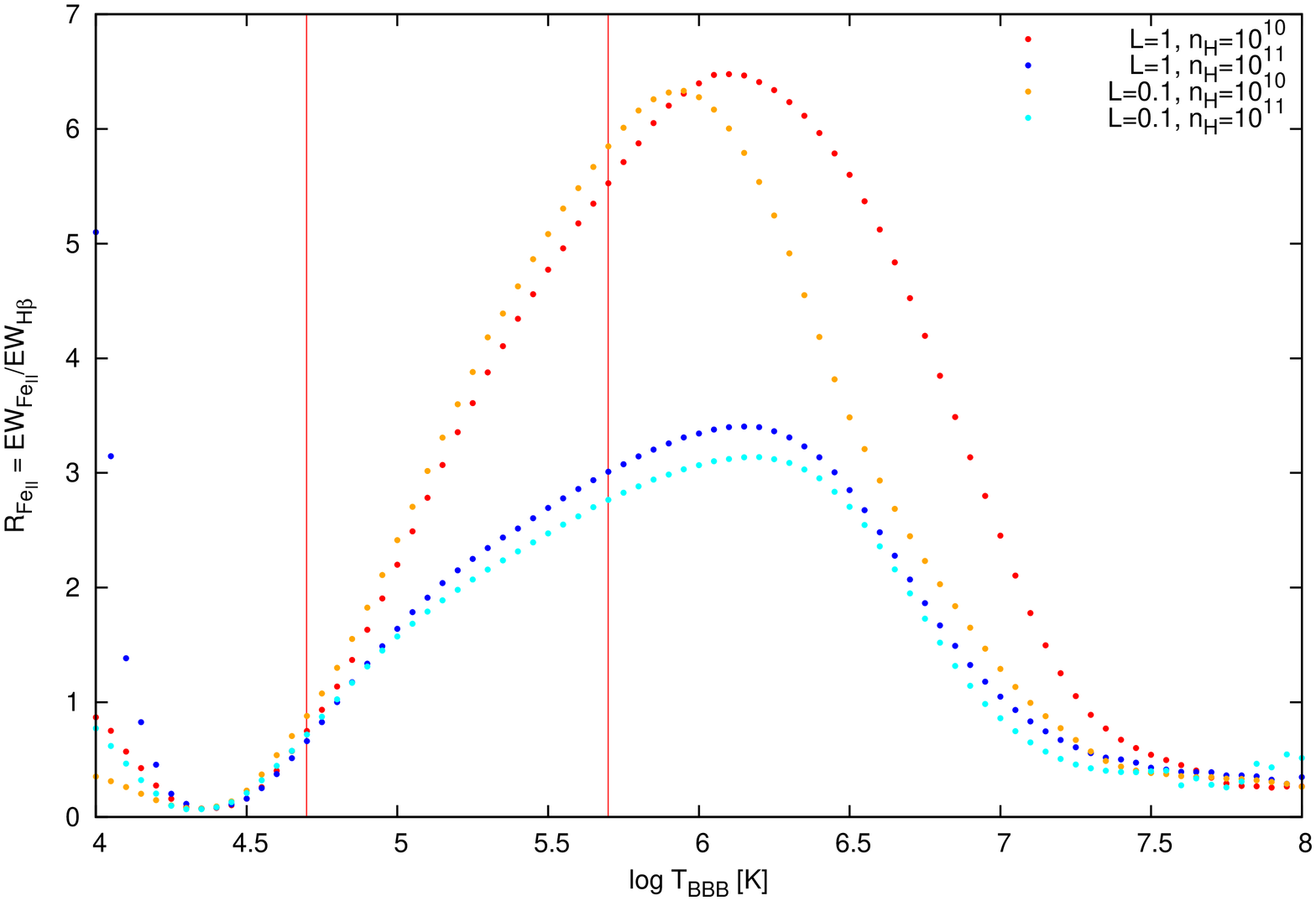}
\label{fig:figure1a}
\end{minipage}
\hspace{0.5cm}
\begin{minipage}[b]{0.45\linewidth}
\centering
\includegraphics[width=0.75\textwidth, angle=270]{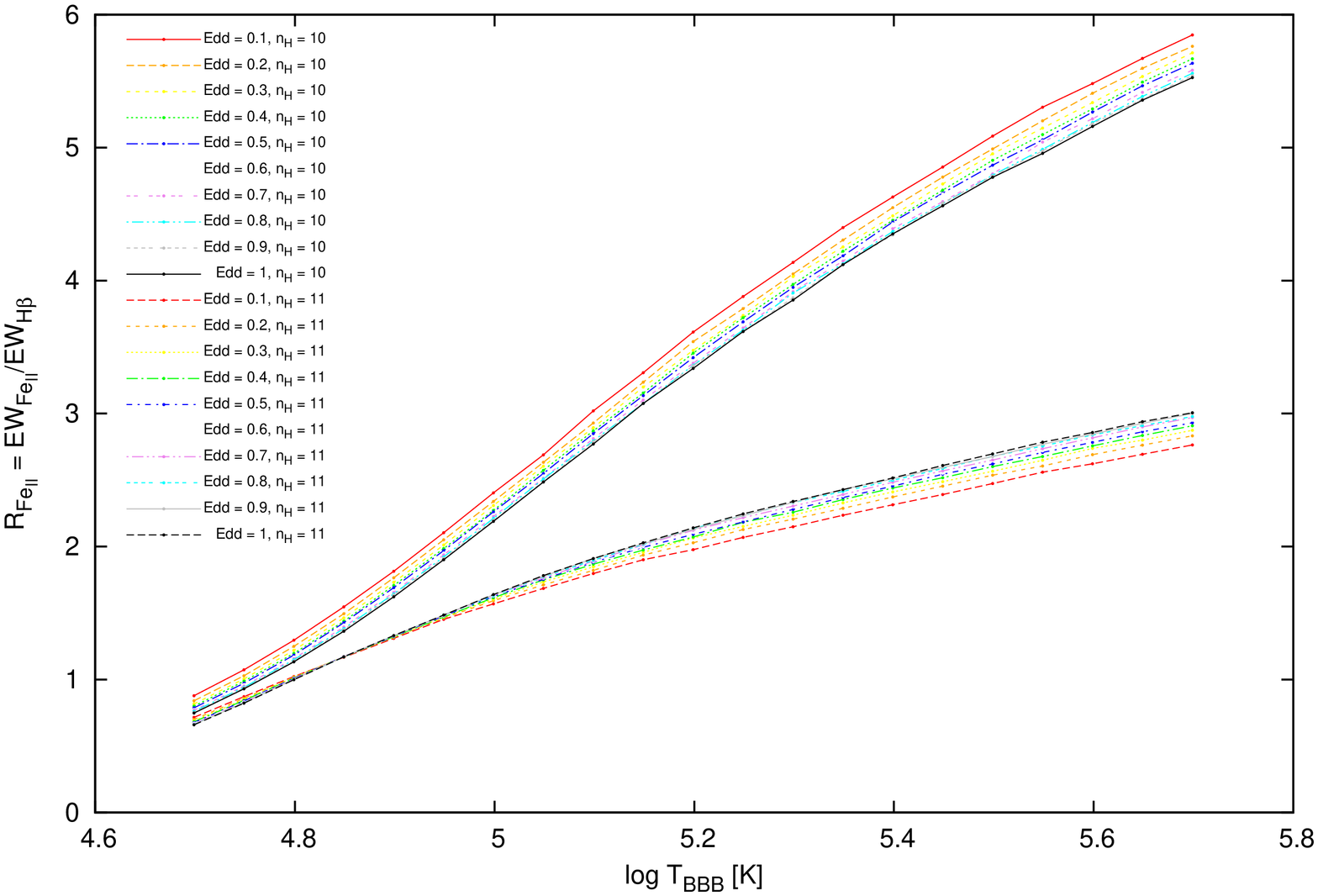}
\label{fig:figure1b}
\end{minipage}
\caption{(a) Comparing $\mathrm{R_{FeII} - T_{BBB}}$ trends between two cases of Eddington ratio $\left ( \mathrm{\frac{L_{bol}}{L_{Edd}} = 0.1\; \&\; 1} \right )$. The vertical red lines specify the maximum disk temperatures for which the derived black hole masses lie in the range [$\mathrm{6.06\times 10^5\;M_{\odot}, 6.06\times 10^9\;M_{\odot}}$]; (b) Comparing $\mathrm{R_{FeII} - T_{BBB}}$ trends for 10 consecutive cases of change in Eddington ratio in the range [0.1, 1] within the considered maximum disk temperature range [$5\times 10^4\;\mathrm{K} - 5\times10^5\;\mathrm{K}$]. For all of the cases, we vary the mean hydrogen density ($\mathrm{n_H = 10^{10}\; cm^{-3}\; \&\; 10^{11}\; cm^{-3}}$)}
\end{figure}

In Fig. 1(b), we zoom the simulations into the considered range of $\mathrm{T_{BBB}}$ and show the $\mathrm{R_{FeII} - T_{BBB}}$ trends by varying the Eddington ratio within the range [0.1,1] with a step size of 0.1. We test these cases again by varying the $\mathrm{n_H}$ as in Fig. 1(a). Although there is not a strong dependence of $\mathrm{R_{FeII}}$ on the change in Eddington ratio, for the lower $\mathrm{n_H}$ i.e. $\mathrm{10^{10}\;cm^{-3}}$, the trends rise uniformly with the trend corresponding to the lowest Eddington ratio case has the maximum value for $\mathrm{R_{FeII}} = 5.847$. But in the higher $\mathrm{n_H}$ case, the trends converge at log $\mathrm{T_{BBB}}$ = 4.899, but then diverge. After the divergence the slope of trend corresponding to the highest Eddington ratio case (=1.0) is the highest and  has the maximum value for $\mathrm{R_{FeII}} = 3.007$. 
\par 
We also include microturbulence to test its effect on the existing $\mathrm{R_{FeII} - T_{BBB}}$ trends. We consider five distinct scenarios ($\mathrm{v_{turb} = 10,20,30,50,100\;km/s}$) and plot the subsequent $\mathrm{R_{FeII} - T_{BBB}}$ trends obtained alongside the existing trend (without microturbulence). We currently limit to $\mathrm{n_H = 10^{11}\; cm^{-3},\; N_H = 10^{24}\; cm^{-2}}$. The maximum of the peak value of $\mathrm{R_{FeII}}$ is obtained for the original case without any microturbulence. The peak of $\mathrm{R_{FeII}}$ subsequently drops to its lowest value for $\mathrm{v_{turb} = 10\;km/s}$ and rises as the turbulent velocity is increased till $\mathrm{v_{turb} = 100\;km/s}$ falling short of the original peak $\mathrm{R_{FeII}}$. The peak values mentioned are that of the maximum value obtained within the temperature range $\mathrm{2.24 \times 10^4\;K\; -\; 10^8\;K}$ as the rise below $\mathrm{2.24 \times 10^4\;K}$ is as explained above due to the effect of $\mathrm{\alpha_{ox}}$. But the temperature corresponding to the peak value of $\mathrm{R_{FeII}}$ shifts towards higher temperatures with increasing turbulent velocity.
\begin{figure}[ht]
\begin{minipage}[b]{0.45\linewidth}
\centering
\includegraphics[width=0.75\textwidth, angle=270]{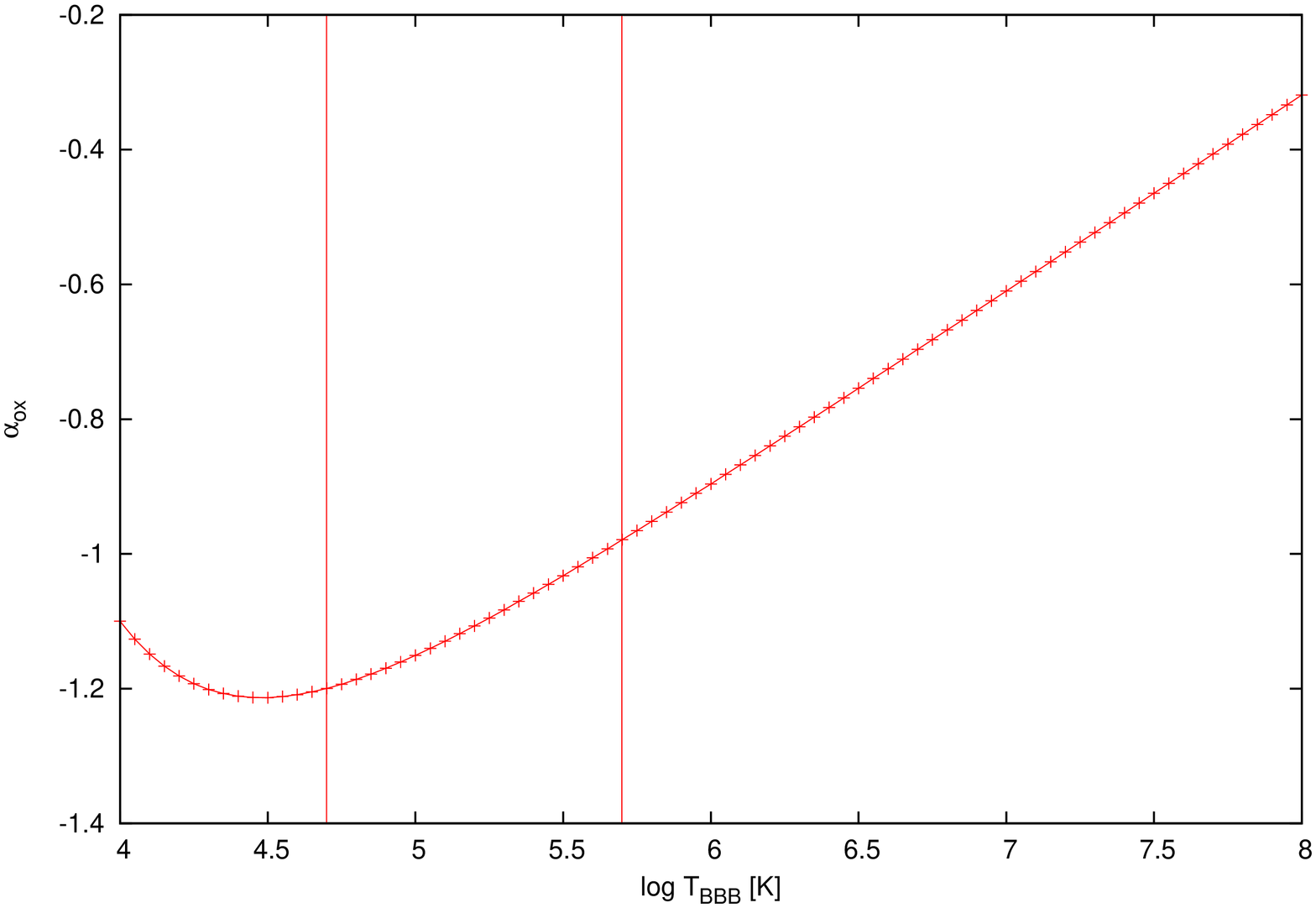}
\label{fig:figure2a}
\end{minipage}
\hspace{0.5cm}
\begin{minipage}[b]{0.45\linewidth}
\centering
\includegraphics[width=0.75\textwidth, angle=270]{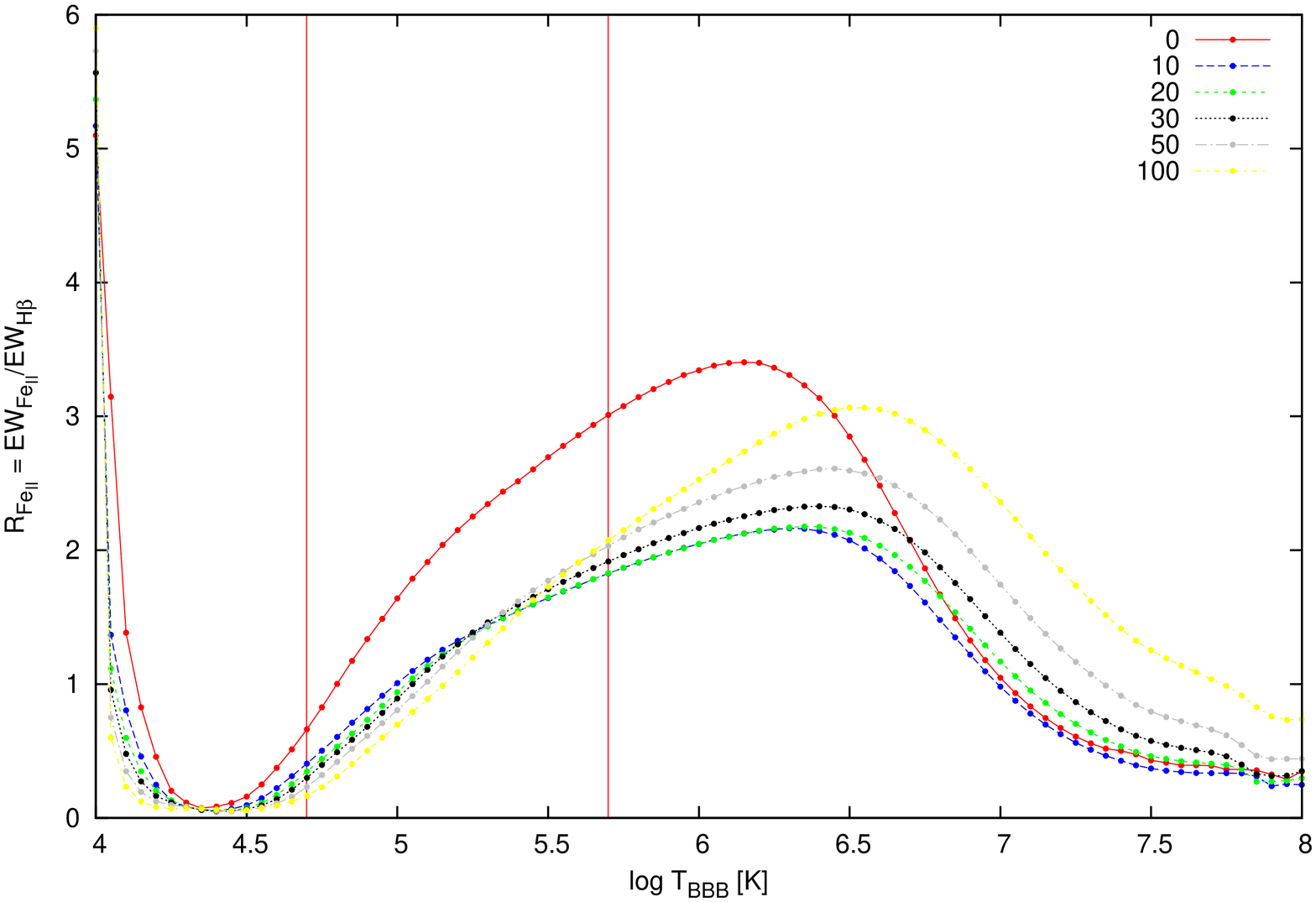}
\label{fig:figure2b}
\end{minipage}
\caption{(a) The changing broad band spectral index, $\mathrm{\alpha_{ox}}$ with increasing disk temperature ($\mathrm{T_{BBB}}$) . The trend is almost linear in the considered temperature range shown by the vertical red lines. (b) Comparing $\mathrm{R_{FeII} - T_{BBB}}$ trends for six different cases of turbulent velocity ($\mathrm{v_{turb} = 10,20,30,50,100\;km/s}$) at $\mathrm{n_H = 10^{11}\; cm^{-3},\; N_H = 10^{24}\; cm^{-2}}$.}
\vspace{0.5cm}
\textbf{Acknowledgement:} The project is supported by Polish grant No. 2015/17/B/ST9/03436/.\\ 
\end{figure}

\bibliographystyle{ptapap}
\bibliography{ptapap}

\begin{thebibliography}{13}
\providecommand{\natexlab}[1]{#1}
\providecommand{\url}[1]{\texttt{#1}}
\providecommand{\urlprefix}{URL }
\providecommand{\eprint}[2][]{\url{#2}}

\bibitem[{{Boroson} \& {Green}(1992)}]{bg92}
{Boroson}, T.~A., {Green}, R.~F., \emph{{The emission-line properties of
  low-redshift quasi-stellar objects}}, \emph{ApJS} \textbf{80}, 109 (1992)

\bibitem[{{Ferland} et~al.(2013)}]{f13}
{Ferland}, G.~J., et~al., \emph{{The 2013 Release of Cloudy}}, \emph{RMxAA}
  \textbf{49}, 137 (2013), \eprint{1302.4485}

\bibitem[{{Kuraszkiewicz} et~al.(2009)}]{kura09}
{Kuraszkiewicz}, J., et~al., \emph{{Principal Component Analysis of the
  Spectral Energy Distribution and Emission Line Properties of Red 2MASS Active
  Galactic Nuclei}}, \emph{ApJ} \textbf{692}, 1180 (2009), \eprint{0810.5714}

\bibitem[{{Lusso} \& {Risaliti}(2017)}]{lusso17}
{Lusso}, E., {Risaliti}, G., \emph{{Quasars as standard candles. I. The
  physical relation between disc and coronal emission}}, \emph{\aap}
  \textbf{602}, A79 (2017), \eprint{1703.05299}

\bibitem[{Panda et~al.(2017a)Panda, Czerny, \& Wildy}]{panda17a}
Panda, S., Czerny, B., Wildy, C., \emph{The Physical Driver of the Optical
  Eigenvector 1 in Quasar Main Sequence}, \emph{Frontiers in Astronomy and
  Space Sciences} \textbf{4}, 33 (2017a),
  \urlprefix\url{https://www.frontiersin.org/article/10.3389/fspas.2017.00033}

\bibitem[{{Panda} et~al.(2017b){Panda}, {Czerny}, {Wildy}, \&
  {{\'S}niegowska}}]{panda17b}
{Panda}, S., {Czerny}, B., {Wildy}, C., {{\'S}niegowska}, M., \emph{{Testing
  the physical driver of Eigenvector 1 in Quasar Main Sequence}}, \emph{ArXiv
  e-prints}  (2017b), \eprint{1712.05176}

\bibitem[{{Shen} \& {Ho}(2014)}]{sh14}
{Shen}, Y., {Ho}, L.~C., \emph{{The diversity of quasars unified by accretion
  and orientation}}, \emph{Nature} \textbf{513}, 210 (2014), \eprint{1409.2887}

\bibitem[{{Sulentic} et~al.(2001){Sulentic}, {Calvani}, \& {Marziani}}]{sul01}
{Sulentic}, J.~W., {Calvani}, M., {Marziani}, P., \emph{{Eigenvector 1: an H-R
  diagram for AGN?}}, \emph{The Messenger} \textbf{104}, 25 (2001)

\bibitem[{{Sulentic} et~al.(2007){Sulentic}, {Dultzin-Hacyan}, \&
  {Marziani}}]{sul07}
{Sulentic}, J.~W., {Dultzin-Hacyan}, D., {Marziani}, P., \emph{{Eigenvector 1:
  Towards AGN Spectroscopic Unification}}, in S.~{Kurtz} (ed.) Revista Mexicana
  de Astronomia y Astrofisica Conference Series, \emph{Revista Mexicana de
  Astronomia y Astrofisica, vol.~27}, volume~28, 83--88 (2007)

\bibitem[{{Sulentic} et~al.(2000){Sulentic}, {Zwitter}, {Marziani}, \&
  {Dultzin-Hacyan}}]{sul00}
{Sulentic}, J.~W., {Zwitter}, T., {Marziani}, P., {Dultzin-Hacyan}, D.,
  \emph{{Eigenvector 1: An Optimal Correlation Space for Active Galactic
  Nuclei}}, \emph{ApJl} \textbf{536}, L5 (2000), \eprint{astro-ph/0005177}

\bibitem[{{Sulentic} et~al.(2002)}]{sul02}
{Sulentic}, J.~W., et~al., \emph{{Average Quasar Spectra in the Context of
  Eigenvector 1}}, \emph{ApJl} \textbf{566}, L71 (2002),
  \eprint{astro-ph/0201362}

\bibitem[{{Sun} \& {Shen}(2015)}]{sun15}
{Sun}, J., {Shen}, Y., \emph{{Dissecting the Quasar Main Sequence: Insight from
  Host Galaxy Properties}}, \emph{ApJl} \textbf{804}, L15 (2015),
  \eprint{1503.08364}

\bibitem[{{Yip} et~al.(2004)}]{yip04}
{Yip}, C.~W., et~al., \emph{{Spectral Classification of Quasars in the Sloan
  Digital Sky Survey: Eigenspectra, Redshift, and Luminosity Effects}},
  \emph{AJ} \textbf{128}, 2603 (2004), \eprint{astro-ph/0408578}

\end{thebibliography}
\end{document}